# High-Performance Thermoelectric Properties of Half-Heusler CoHfSi: A First-Principles Study with Temperature-Dependent Relaxation Time

Sadhana Matth, S. Pandey, Himanshu Pandey*

*Condensed Matter & Low-Dimensional Systems Laboratory, Department of Physics, Sardar Vallabhbhai National Institute of Technology, Surat-395007, Gujarat, India*

*hp@phy.svnit.ac.in

## Abstract

In the ongoing search for innovative thermoelectric (TE) materials with superior TE performance globally, we aim to investigate the possible use of *half*-Heusler alloy CoHfSi in TE applications. We analyzed the structure stability, thermodynamic inertia and electrical and thermal transport properties using density functional formalism and semi-classical Boltzmann transport theory. Positive phonon frequencies confirm this alloy's dynamical stability, and the Born-Huang stability criterion is also satisfied, confirming the robust mechanical stability. A large Seebeck voltage of more than 150 μV/K is estimated, an essential and typical requirement for improved heat-to-electricity conversion efficiency. This Seebeck voltage can be further increased by an order of magnitude with suitable doping. The PHONO3PY algorithm and Slack's model are used to compare the lattice thermal conductivity. The latter method gives more values than the former algorithm. Despite the commonly used constant relaxation time approximation to estimate the TE performance, we adopt the temperature-dependent relaxation time and found a clear drop in figure-of merit ($zT$) from those estimated without considering the lattice thermal conductivity and relaxation time both, still, the $zT$ values are remarkably more than two, for the temperatures above 500 K, which is a striking numeral in the field of TE materials.

## 1. Introduction

The pressing issues of global warming and energy generation require us to accelerate our pace of energy production while minimizing $CO_2$ emissions and reducing reliance on finite fossil fuel resources. Developing advanced renewable energy technologies has become a paramount global priority to meet the energy demands of the growing world population. Although renewable sources such as wind, solar, tidal, biomass, and geothermal energy offer promising solutions, their intermittent nature and dispersed availability pose challenges compared to the centralized power plants currently dominating the electrical energy landscape. Therefore, it is crucial to explore strategies that enhance the reliability and accessibility of these renewable energy sources. Nevertheless, a considerable proportion of energy is expended as heat through the combustion of fossil fuels and other traditional and renewable energy sources. If this energy could be used and converted into electricity with optimal effectiveness and efficiency, it would serve as a promising prospective electrical power source. Intensive research into developing and enhancing thermoelectric materials has been motivated by the increasing demand for clean energy solutions and the necessity to improve the efficiency of current energy systems [1] [2] [3] [4] [5].

Thermoelectric (TE) materials have received much attention due to the growing demand for renewable energy resources to convert waste heat from various industrial machines to usable electricity using the Seebeck effect. However, the low conversion efficiency of these materials has been a critical hindrance to replacing old power generation technologies [6] [7]. The efficiency of TE materials can be calculated using a dimensionless quantity known as a figure of merit ($zT$), which takes into account the Seebeck coefficient ($S$), electrical conductivity ($\sigma$), and thermal conductivity ($\kappa$) consisting of electronic ($\kappa_e$) as well as lattice ($\kappa_l$) parts.

$$zT = \frac{\sigma S^2}{(\kappa_e + \kappa_l)} T \tag{1}$$

Here, $T$ stands for temperature on the absolute scale. Inherent trade-offs between these factors offer a challenge in designing high-performance TE materials. Materials with a higher value of $zT$ provide more conversion efficiency. To maximize $zT$, a balance between the numerator $\sigma S^2$, also known as the power factor, and the denominator $\kappa$ $(= \kappa_e + \kappa_l)$ is needed to maintain.

Researchers have been searching for newer materials to improve $zT$. Various families of materials, such as Clathrates [8], Chalcogenides [9], Skutterudites [10], Heusler alloys [11] [12] [13] [14] [15] (especially cobalt-based Heusler alloys), and many others, are being investigated theoretically as well as experimentally. Due to robustness, low cost, thermodynamic stability, non-toxic [16], and a multi-functional class of materials, Heusler alloys have also been extensively investigated and synthesized for various energy-based applications such as spintronics [17] [18] [19] metal-ion batteries [20] [21] catalyst [22] sensor [23] [24] shape memory [25] [26] energy harvesting [27] [28] and magnetic cooling [29] *etc*. Due to their tunable electrical transport properties via band engineering and rich elemental combinations, they have gained significant attention among the scientific community. An inherent benefit of Heusler alloys in TE applications is their capacity to attain high power factors while preserving comparatively low thermal conductivity. This is mainly attributed to the unique crystal structures of these materials, which may be precisely controlled to increase phonon scattering and decrease lattice thermal conductivity without causing a substantial decrease in electrical conductivity. *Half*-Heusler (HH) alloys (*ABC* type) typically comprise transition metals (*A* and *B*) and main group elements (*C*). The bonding between *A* and *C* atoms mainly creates a semiconducting band gap [30]. These alloys are being investigated to tailor the band gap and the Fermi level ($E_F$) position, which helps to design *p-type* and *n-type* semiconductors from the same alloy with comparable physical properties. This minimizes the lattice mismatch where different *p-type* and *n-type* materials and respective devices can be fabricated using the same parent alloy, and can offer a higher energy conversion. So far, most of the efforts have been concentrated on improving the TE performance of *n-type* HH alloys. However, there are few studies on *p-type* HH alloys, and $zT$ values are still relatively low. Therefore, it is essential and necessary to explore potential *p-type* HH alloys or enhance the performance of existing *p-type* HH alloys [31] [32] [33] [34]. The $zT$ of *p-type* FeNbSb Heusler alloys have been reported around 1.5 at 1200 K [35] [36], whereas for *n-type* XNiSn ($X$ = Ti, Zr, and Hf) alloys, around one at 820 K, and that of *p-type* XCoSb ($X$ = Ti, Zr, and Hf) alloys ~ 1 at 1073 K [37] [38].

Hafnium and silicon-based Heusler alloys have been significantly developed for their TE applications. Hence, we have chosen the CoHfSi HH alloy to boost the *zT* values for this work. Also, before going for any experimental realization of an actual device, it is essential to understand the characteristics, correlation, tunability, and performance of any material through the first-principles calculations that can offer valuable and early predictions. The main objective of this work is to investigate the electronic structure, thermodynamic stability, and the TE properties of CoHfSi. The results obtained from this study indicate that CoHfSi has an apt potential for TE applications.

## 2. Computational Details

Firstly, we optimized the crystal structure of CoHfSi in the *face-centred* cubic phase (space group $F\bar{4}3m$, 216) within generalized gradient approximation [39] by estimating the ground state energy using the first-principle calculation as implemented in the *Quantum Espresso* package [40]. Ultrasoft pseudo-potentials were used to study the interaction between core and valence electrons. The *cut-off* corresponding to kinetic energy for wave functions and charge densities was set to be 100 and 900 Ry, respectively. In the Brillouin zone, a *k*-point mesh of 12 × 12 × 12 and a much denser mesh of 40 × 40 × 40 were considered, with a convergence threshold criterion of $10^{-6}$ Ry for optimization and electronic structure calculations, respectively. A semi-classical code based on Boltzmann transport formalism, *BoltzTrap2* [41], was used to calculate the TE properties for CoHfSi HH alloy. The values of *S*, $\sigma$, and $\kappa_e$, were calculated using Eqs. (2) and (3) via the *BoltzTrap2* utility [42].

$$S_{\alpha\beta}(T,\mu) = \frac{1}{eT\Omega\sigma_{\alpha\beta}} \int \sigma_{\alpha\beta}(\varepsilon)(\varepsilon-\mu)\left[-\frac{\partial f_0(T,\varepsilon,\mu)}{\partial\varepsilon}\right] d\varepsilon \tag{2}$$

$$\sigma_{\alpha\beta}(T,\mu) = \frac{1}{\Omega} \int \sigma_{\alpha\beta}(\varepsilon)\left[-\frac{\partial f_0(T,\varepsilon,\mu)}{\partial\varepsilon}\right] d\varepsilon \tag{3}$$

As evident from Eq. (1), the *zT* values also depend on $\kappa_l$, its contribution needs to be considered. Here, we have employed Slack's model [43] to estimate $\kappa_l$ as given by Eq. (4).

$$\kappa_l = 2.43\times10^{-8}\left[1-\frac{0.514}{\gamma}+\frac{0.228}{\gamma^2}\right]^{-1}\left(\frac{\bar{M}n^{1/3}\delta\theta_D^3}{T\gamma^2}\right) \tag{4}$$

Here, $\bar{M}$, *n*, $\theta_D$, and $\delta$ represent the average atomic mass of CoHfSi, the total number of atoms in the unit cell, the Debye temperature associated with acoustic phonon modes, and the volume of the unit cell per atom, respectively. Also, $\gamma$ is the temperature-dependent Grüneisen parameter, calculated from the Poisson ratio ($\nu$), $\gamma = \frac{3}{2}\left[\frac{1+\nu}{2-3\nu}\right]$. The temperature-dependent values of $\kappa_l$, as estimated from Eq. (4) are compared to those calculated by the *Phono3py* package [44]. Here, the relaxation time ($\tau$), another crucial parameter affecting the charge carrier transport, is also included while estimating *zT* from $\sigma$ and $\kappa_e$ as these two quantities are generally calculated in terms of $\tau$.

## 3. Results and Discussion

### 3.1 Formation Energy and Phonon Dispersion

CoHfSi HH exhibits three distinct structural configurations depending on the Wyckoff positions of the constituting atoms in the unit cell. A rigorous investigation of the structure optimization has been presented earlier [45]. To check the stability of this alloy, formation energy is calculated as $E_{form} = E_{CoHfSi} - (E_{Co} + E_{Hf} + E_{Si})$. Here, $E_i$'s represents the ground state energy of the alloy in its unit cell and isolated atoms. A negative value of $E_{form}$ is estimated, representing the thermodynamic stability of CoHfSi alloy. The lattice parameters, formation energy ($E_{form}$), effective mass ($m^*$), and band gap ($E_g$) values are summarized in Table 1.

Table 1: Calculated lattice parameter ($a$), the effective mass of holes/electrons ($m_h$*/$m_e$*, in the units of $m_e$) computed at VBM and CBM, formation energy ($E_{form}$), and band gap ($E_g$) for CoHfSi alloy.

| $a$ (Å) | $m_h$* | $m_e$* | $E_{form}$ (eV) | $E_g$ (eV) |
|---|---|---|---|---|
| 5.770 | 0.300 | 1.208 | -21.375 | 1.132 |

Phonon calculation is another tool used to check the dynamic stability of any material. The phonon dispersion curve and phonon density of states (DOSs) calculated for the CoHfSi HH alloy are presented in Figs. 1(a) and 1(b), respectively. The absence of any imaginary and negative phonon frequency along the high symmetry *k-points* confirms the stability. The lower two transverse acoustic (TA) bands (≈ 70-90 $cm^{-1}$) share the same energy along the path $L \rightarrow \Gamma \rightarrow X$. Above this band, a longitudinal acoustic (LA) band (≈ 110 $cm^{-1}$) occurs, which exhibits a degeneracy at the $\Gamma$-point with the TA bands. In the high-frequency regime (≈ 200–280 $cm^{-1}$), four transverse optical (TO) and two longitudinal optical (LO) bands can be seen. These also show a similar kind of degeneracy as the acoustic bands. The separation between the optical and acoustic bands plays a vital role for $\kappa_l$: The smaller the gap between these branches, the lower $\kappa_l$, helps to attain a higher $zT$ value [13]. From the phonon DOSs [Fig. 1(b)], the Co atom dominates at a lower frequency range, whereas Si and Hf dominate in the high-frequency range.

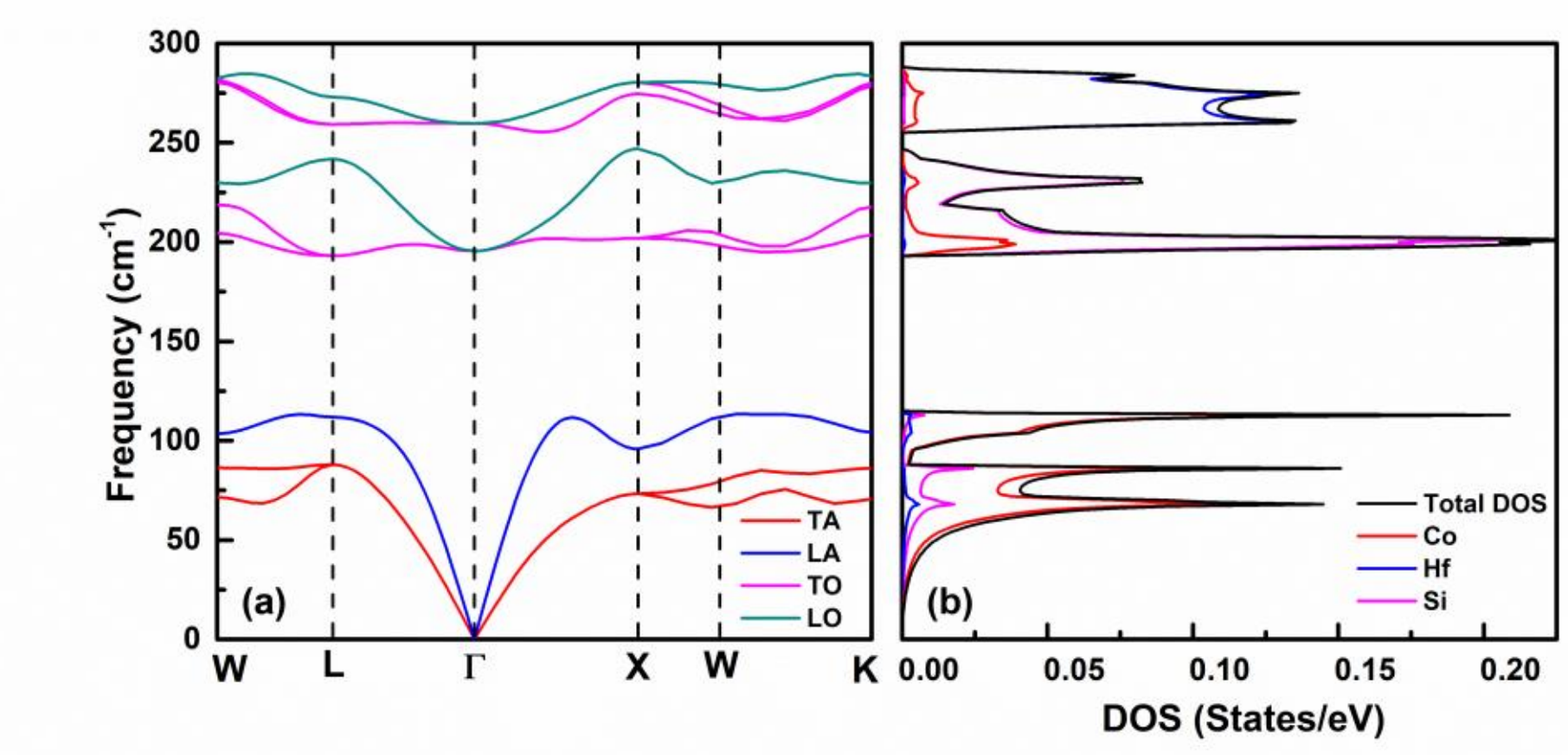

Figure 1: (a) Phonon band structure and (b) phonon density of states calculated for CoHfSi unit cell.

### 3.2 Elastic and Thermodynamic Properties

Apart from the materials' stability, their mechanical properties are equally important in investigating their feasibility for robust application, such as in any proposed real device.

Several parameters, including the interatomic distance, bond length, and arrangement of atoms in the crystal, influence the stability of a material and are examined by their elastic properties, providing information about the mechanical response to an external force. Under the Voigt approximation, a cubic crystal's three independent elastic coefficients ($C_{11}$, $C_{12}$, and $C_{44}$), are calculated using the *thermo_pw* package [31] [46]. These coefficients indicate materials' resistance to linear compression and transverse expansion under the application of stress and are summarized in Table 2. If these three elastic constants adhere to the Born-Huang stability criterion [47], as given by Eq. (5), then the material is considered stable.

$$C_{11}>0,\ C_{44}>0,\ (C_{11}-C_{12})>0,\ (C_{11}+2C_{12})>0 \tag{5}$$

Table 2: Calculated elastic constants $C_{11}$, $C_{12}$, $C_{13}$ and derived parameters bulk modulus ($B$), the Young's modulus ($Y$), shear modulus ($G$), the Cauchy pressure ($CP$) [all in GPa]; the Debye temperature ($\theta_D$) in K; dimensionless Poisson ratio ($v$), and the Pugh ratio ($B/G$).

| $C_{11}$ | $C_{12}$ | $C_{44}$ | $B$ | $Y$ | $G$ | $CP$ | $\theta_D$ | $v$ | $B/G$ |
|---|---|---|---|---|---|---|---|---|---|
| 242.23 | 98.01 | 45.19 | 146.08 | 148.87 | 55.96 | 43.82 | 321.67 | 0.33 | 2.61 |

Additional elastic properties of the material can be determined from $C_{11}$, $C_{12}$, and $C_{44}$. For example. Bulk modulus ($B$): determines how much a material resists its volume change; the Young's modulus ($Y$): determines the materials' capacity to resist change in length when it is stretched or compressed in one direction; shear modulus ($G$): suggests about plastic deformation; the Poisson's ratio ($v$), Pugh ratio ($B/G$), and Cauchy pressure ($CP = C_{12}-C_{44}$): determine whether the material is ductile or brittle. $v < 0.28$, $CP > 0$, $B/G > 1.75$ indicate that the material is ductile in nature; otherwise, it is brittle [31]. A positive value of $CP$ infers an ionic bonding between the atoms; otherwise, it will be mainly covalent. Another thermodynamic parameter, the Debye temperature ($\theta_D$), provides information about the lattice vibrations. The estimated values for CoHfSi HH alloy are tabulated in Table 2.

To get more insight into thermal stability and thermodynamic characteristics, the quasi-harmonic Debye model is applied to predict the temperature dependence of properties such as heat capacity, entropy, vibrational energy, and vibrational free energy, which are calculated in a temperature range of 100–800 K and shown in Fig. 2(a-d). Increasing temperature makes a more rapid atomic motion possible due to translational, rotational, and vibrational degrees of freedom, resulting in increased kinetic energy [Fig. 2(a)]. Figure 2(b) depicts the variation of entropy with temperature, which is found to increase with the increase in temperature as it measures the system's internal energy. The variation of the Debye heat capacity with temperature is plotted in Fig. 2(c) and found to vary linearly up to 100 K and above 600 K, where it becomes independent of temperature and approaches the Dulong-Petit limit of 75 J/(K mol). The Debye vibrational free energy is found to decrease with the increase in temperature, as given in Fig. 2(d). Material with more negative vibrational energy becomes more thermodynamically stable and has a better thermal response [48].

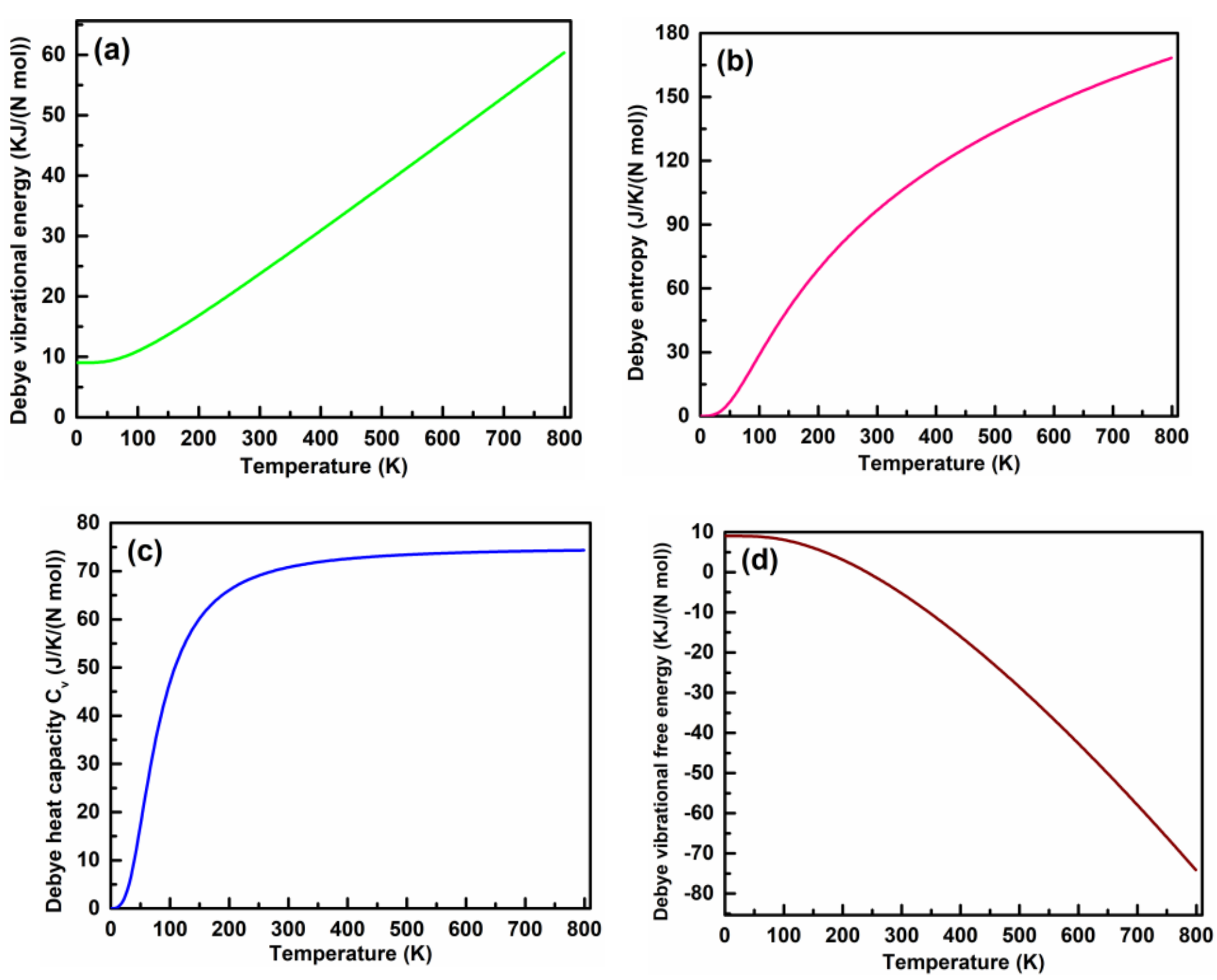

Figure 2: Temperature-dependent variation of a) Debye vibrational energy, b) Debye entropy, c) Debye heat capacity, and d) Debye vibrational free energy.

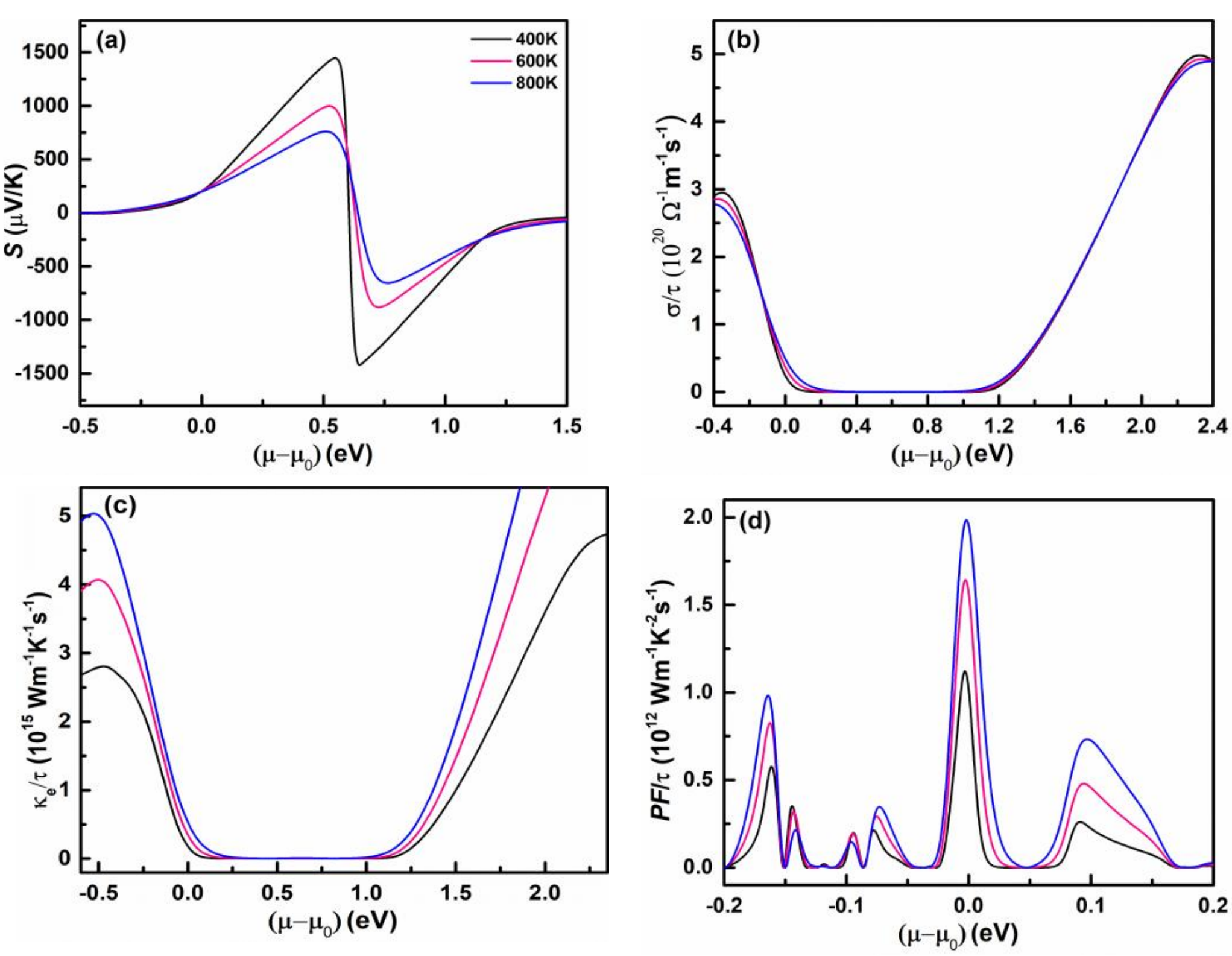

Figure 3: Variation of various thermoelectric parameters a) $S$, b) $\sigma/\tau$, c) $\kappa_e/\tau$, d) $PF/\tau$ with the position of relative chemical potential ($\mu$) at 300 K, 500 K, 700K and 800 K for the CoHfSi.

### 3.3 Thermoelectric Properties

The Seebeck coefficient is plotted as a function of relative chemical potential ($\mu$) to the Fermi level ($E_F$) at 300 K, 500 K, 700 K, and 800 K, as given in Fig. 4 (a). The peak value of $S$ at 400 K for *n-type* is 1450 μV/K; however, with the increase in temperature, this peak value decreases and attains 712 μV/K at 800 K. This is due to an increase in thermal energy. The $\sigma/\tau$ values are plotted in Fig. 4(b) as a function of ($E_F$ -$\mu$). It is observed as zero in mid values of ($E_F$ -$\mu$), a typical semiconducting material behaviour. However, it can be seen that where the $S$ value has its maximum value, the $\sigma/\tau$ is virtually approaching zero. As we go away from this range, $\sigma/\tau$ increases. This is caused by a low density of charge carriers, which lowers conductivity. Beyond the Fermi level, two peaks exist, one in the *p-type* region and the second in the *n-type* region. In the *n-type* region, maximum peaks of $\sigma/\tau$ are around $4.9\times10^{20}$ $\Omega^{-1}m^{-1}s^{-1}$, whereas in the *p-type* region, the highest values are close to $3\times10^{20}$ $\Omega^{-1}m^{-1}s^{-1}$. It clearly represents that *n-type* doping will dominate over *p-type* doping for the CoHfSi compound due to the lighter effective mass of electrons in the conduction band. Another important thermoelectric parameter, $\kappa_e/\tau$, is plotted in Fig. 4(c) as a function of ($E_F$ -$\mu$) for different temperatures. For the *n-type*, these values are higher than those for the *p-type* for all temperatures. The $\kappa_e/\tau$ value decreases with temperature for both *p-type* and *n-type*, reaching $7\times10^{15}$ $Wm^{-1}K^{-1}s^{-1}$ and $5\times10^{15}$ $Wm^{-1}K^{-1}s^{-1}$ at 800 K. The plots of $\kappa_e/\tau$ and $\sigma/\tau$ are nearly identical as these quantities are interconnected through the Wiedemann-Franz law. The variation of $PF/\tau$ with $\mu$-$\mu_0$ estimated for different temperatures is shown in Fig. 4 (d). A maximum value around $2\times10^{12}$ $Wm^{-1}K^{-2}s^{-1}$ for $PF/\tau$ is found near the Fermi level for 800 K, and this maximum value decreases very sharply as we move away from $E_F$ on either side for all the temperatures [37].

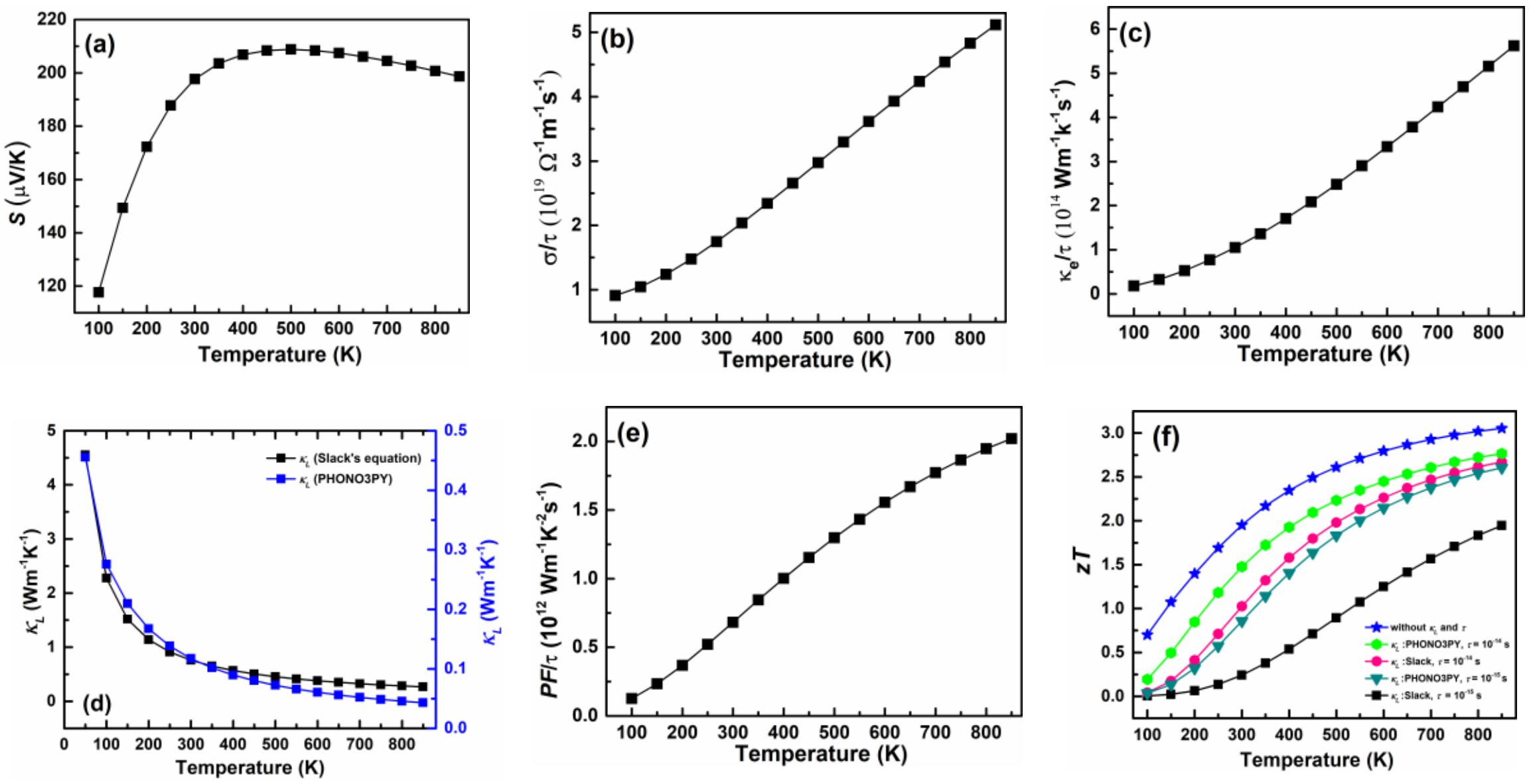

Figure 4: Variation of a) Seebeck coefficient, b) electrical conductivity per relaxation time, c) electronic thermal conductivity per relaxation time, d) lattice thermal conductivity, e) power factor per relaxation time, f) $zT$ with temperature for the CoHfSi unit cell.

Figure 4(a) demonstrates the relationship between *S* and *T*, exhibiting a linear increase with temperature, reaching a maximum and then starting to decrease gradually. This high-temperature reduction in *S* can be attributed to (i) an increase in minority carriers at elevated temperatures, which further increases the bipolar transport; (ii) a decrease in unoccupied states above the Fermi level at high temperatures; and (iii) an increase in thermally excited charge carriers. The combined effect of all these processes led to lower *S* values [49]. It should also be noted that the positive *S* values across the entire temperature range indicate that holes are mainly responsible for the conduction in this material. For a good thermoelectric material, the value of *S* lies in the range of 150-200 $\mu V/K$. The substantially higher number of states in the valence band (near the Fermi level) than in the conduction band may be one of the reasons for this higher *S* [50] [51]. Figure 3(b) illustrates the temperature variation of $\sigma/\tau$, which increases with temperature, indicating a semiconducting nature in CoHfSi. The increased electrical conductivity results from reduced scattering at elevated temperatures, illustrating the material's degenerate semiconductor characteristics [38] [52]. The temperature-dependent values of $\kappa_e/\tau$ increases with increasing temperature, as shown in Fig. 3(c), mainly due to the increased charge carriers at elevated temperatures and an increase in phonon diffusion [51], which improves the electronic thermal conduction. An identical behaviour is observed for both quantities (electronic thermal conductivity and electrical conductivity), indicating a direct proportionality between them.

Since the TE efficiency depends on the total thermal conductivity, the lattice thermal conductivity ($\kappa_l$) is estimated using the PHONO3PY algorithm and Slack's model [Eq. (4)], these two methods are generally implemented to estimate $\kappa_l$. Figure 4(d) illustrates the variation of $\kappa_l$ with temperature. Low values of $\kappa_l$ are found in the 0.5 – 1.0 $Wm^{-1}K^{-1}$ range for temperatures 300 K and above, possibly due to lower phonon group velocity and phonon–phonon lifetimes. A variation of almost one order can be seen; higher values are found by Slack's equation, even after considering the temperature-dependence of Grüneisen parameter ($\gamma$). Due to a lack of experimental data, we cannot comment on the over/under estimation of the respective methods while estimating $\kappa_l$. Apart from the magnitude, the temperature dependence is similar. The $\kappa_l$ values are higher at lower temperatures due to the increased mean free path, which allows greater heat transfer by phonons, resulting in higher lattice thermal conductivity. At elevated temperatures, phonon-phonon scattering prevails, resulting in diminished mean free path, reducing phonons' thermal conductivity. A low value of $\kappa_l$ is necessary to attain higher *zT* values [53]. The thermoelectric power factor ($PF$) represents the combined influence of the *S* and $\sigma/\tau$, and plotted as $PF/\tau$ in Fig. 4(e). A higher $PF$ value indicates that a greater amount of current and voltage is generated. The TE materials with higher $PF$ are in demand to convert waste heat into practical work efficiently Figure 4(f) shows the variation of *zT* with temperatures ranging from 100 K to 850 K. A comparison with the *zT* values estimated without including $\kappa_l$ and $\tau$ values are also given, as in the literature, we mostly encounter the estimation of *zT* for different TE materials, like this. Hence, to get a more reliable prediction of TE performance, first, we include the contribution of $\kappa_l$ (estimated from both PHONO3PY and Slack's method) and consider fixed values of $\tau$. Here, the $\kappa_l$ values from Fig. 4(d) and $\tau \approx 10^{-14}$–$10^{-15}$ *s* are taken to estimate the *zT*. The *zT* values are reduced compared to those estimated without considering $\kappa_l$ and $\tau$. An increase in *zT* with increasing temperature is found. From Fig. 4(e), it is evident that the Slack model apprehends the *zT* values much more in contrast with those computed by the PHONO3PY. Also, a faster relaxation further improves the thermoelectric performance by increasing the *zT* values. Assuming $\tau \approx 10^{-14}$ s, the *zT* values are still more than one, in the range of 1.5 – 2.5 for the temperatures 500 K and above.

But the parameter $\tau$, which represents the various types of scattering mechanisms, *e.g.*, electron-phonon scattering, impurity scattering, is a temperature-dependent quantity and also TE properties $\sigma$, $\kappa_e$, and $\kappa_l$ also depend on $\tau$. So, we consider going beyond the constant relaxation time approximation (CRTA). Most of the reported TE theoretical investigations are based on CRTA with $\tau \approx 10^{-14}$–$10^{-15}$ *s*. In the present work, we went beyond-CRTA and estimated the temperature-dependence of $\tau$, again using Slack's model [43] as given by $\tau = \frac{2\sqrt{2\pi}C\hbar^4}{3\left(k_B m_d^*\right)^{3/2}E^2}\frac{1}{T^{3/2}}$ and for shown in Fig. 5(a) for electrons. Here, apart from the universal physical constants, $C$, $m_d^*$, and $E$ represent the elastic constant, effective mass, and deformation potential, respectively. A sharp decrease in $\tau$ is found at low temperatures due to increased collision frequency between the charge carriers and kinetic energy [14], and then a plateau is obtained after 500 K. The $\tau$ values are estimated to be around $10^{-12}$ s, possibly due to the assumption of a 'perfect crystal' while performing the DFT investigations. In Fig. 5(b), we compare the *zT* by considering the temperature-dependent $\tau$ and $\kappa_l$ obtained from PHONO3PY and Slack's model [Eq. (4)]. These two parameters play a crucial role in realistically modelling transport properties, and their inclusion leads to a more accurate and conservative assessment of material performance. The *zT* values computed with $\kappa_l$ and $\tau$ estimated from Slack's method are nearly the same as those estimated without considering $\kappa_l$ and $\tau$. Hence, it is now clear that Slack's method somewhat overestimates the *zT* values. On the other hand, while considering the $\kappa_l$ values from the PHONO3PY and $\tau$ from Slack's method, the *zT* values are decreased, *e.g.,* a reduction from approximately 3.1 to 2.8 at 800 K and 2.0 to 1.6 at 300 K. Despite this reduction, it is remarkable that the *zT* values at elevated temperatures still exceed 2.5, which is a significant threshold in the domain of TE materials. Surpassing this value indicates a material with strong potential for efficient thermal-to-electrical energy conversion. Achieving *zT* > 2.5, even after accounting for realistic scattering mechanisms and heat dissipation through the lattice, highlights the intrinsic efficiency of the material.

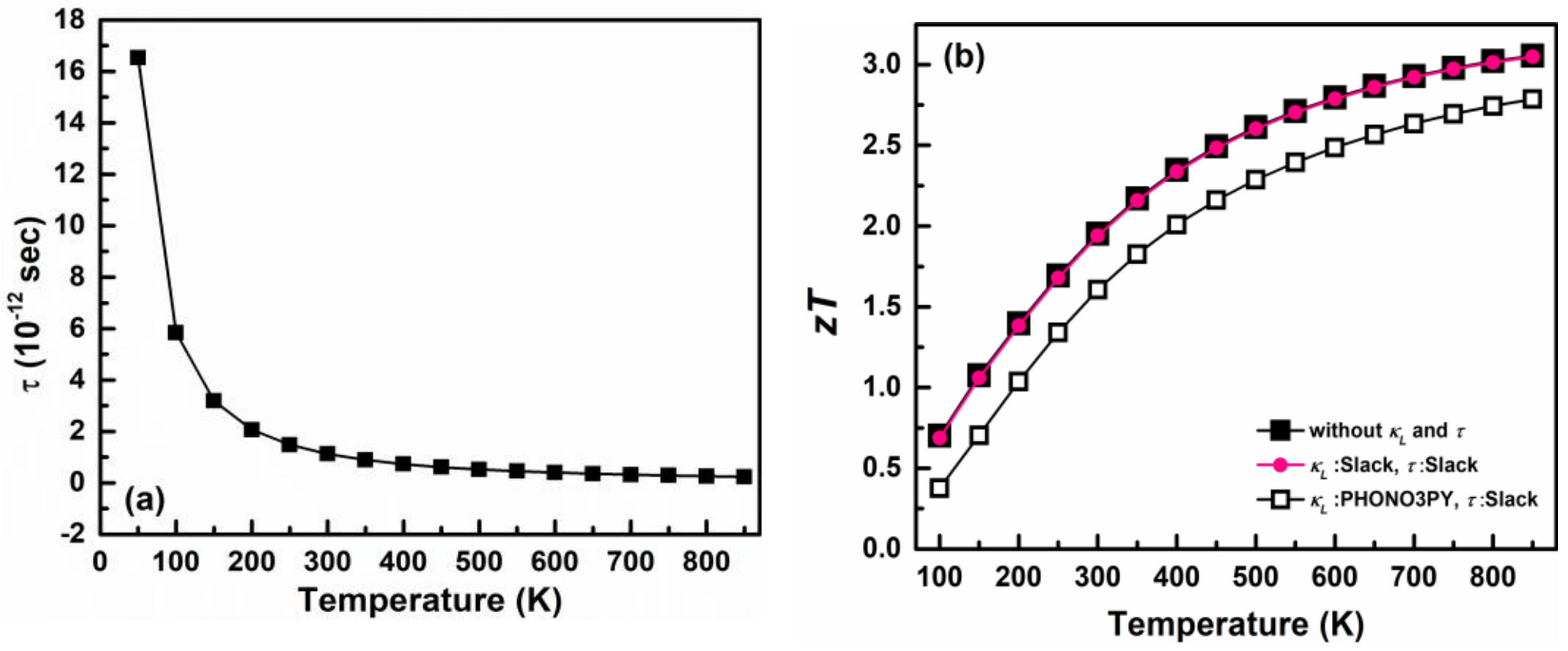

**Figure 5:** Variation of (a) relaxation time with temperature (estimated from Slack's method) and (b) *zT* with temperature to compare Slack and PHONO3PY methods.

## Conclusion

Based on the Kohn-Sham density functional theory, we have investigated the stability and thermoelectric properties of semiconducting *half*-Heusler alloy CoHfSi. A large Seebeck voltage is estimated to be around 150–200 μV/K which can be further increased by an order of magnitude with slight doping. The lattice thermal conductivity is computed by applying PHONO3PY and Slack's model. From the detailed analysis, we found that Slack's method somewhat overestimates the $zT$ values. Also, $zT > 2.5$ is obtained for high temperatures even after considering temperature-dependent relaxation time. This outstanding high-temperature performance positions CoHfSi among a selective group of high-performing TE materials. Materials with such characteristics are highly desirable for various energy applications, including power generation from waste heat and solid-state cooling technologies. The ability of this mechanically robust CoHfSi to maintain superior $zT$ values under more physically rigorous conditions underlines its potential as a reliable and efficient candidate for next-generation TE devices.

## Acknowledgment

This research is supported by the research grant from the Institute Seed Research Grant 2021–22/DOP/04.